\documentclass{ws-procs975x65}

\begin{document}

\title{\uppercase{Shape Dynamics and Gauge-Gravity Duality}}
\author{\uppercase{Henrique Gomes$^1$, Tim A. Koslowski$^2$}}
\address{$^1$ University of California at Davis, One Shields Avenue Davis, CA, 95616, USA\\$^2$ University of New Brunswick, Fredericton, NB, E3B 5A3, Canada\\ email: $^1$\texttt{gomes.ha@gmail.com}, $^2$\texttt{t.a.koslowski@gmail.com}}

\bodymatter

\begin{abstract}
  The dynamics of gravity can be described by two different systems. The first is the familiar spacetime picture of General Relativity, the other is the conformal picture of Shape Dynamics. We argue that the bulk equivalence of General Relativity and Shape Dynamics is a natural setting to discuss familiar bulk/boundary dualities. We discuss consequences of the Shape Dynamics description of gravity as well as the issue why the bulk equivalence is not explicitly seen in the General Relativity description of gravity.
\end{abstract}

\section{Introduction}

The dynamics of pure gravity can be described either as a generally covariant dynamics of the spacetime metric (General Relativity description) or as a spatially covariant and Weyl invariant dynamics of the spatial metric (Shape Dynamics description). Locally, the two descriptions are indistinguishable\cite{SD-refs}. The ADM formulation of General Relativity uses the spatial metric $g_{ab}$ and its canonically conjugate momentum density $\pi^{ab}$ as fundamental variables and is completely described by the first class constraints
\begin{equation}
 \begin{array}{rcl}
   S(N)&=&\int_\Sigma d^3x\,N\,\left(\frac{\pi^{ab}(g_{ac}g_{bd}-\frac 1 2 g_{ab}g_{cd})\pi^{cd}}{\sqrt{|g|}}-(R[g]-2\Lambda)\sqrt{|g|}\right),\\
   ~&~&~\\
   H(v)&=&\int_\Sigma d^3x\,\pi^{ab}(\mathcal L_v g)_{ab},
 \end{array}
\end{equation}
where we consider a compact Cahuchy surface $\Sigma$ without boundary. The constraints $H(v)$ generate spatial diffeomorphisms, while the constraints $S(N)$ generate on-shell refoliations. The restriction of the $S(N)$ to those that preserve a particular foliation generate time evolution within this foliation, meaning that the $S(N)$ entangle constraints and evolution generators. Shape Dynamics shares the spatial diffeomorphism constraints $H(\xi)$ with General Relativity, but replaces the ADM Hamilton constraints with spatial conformal constraints and Hamiltonian: 
\begin{equation}
  \begin{array}{rcl}
    Q(\rho)&=&\int_\Sigma d^3x\,\rho \,(g_{ab}\pi^{ab}-\frac 2 3 \tau\sqrt{|g|}\,),\\
     ~&~&~\\
    H_{SD}&=&\int_\Sigma d^3x \sqrt{|g|} \,\Omega_o^6[g,\pi],
  \end{array}
\end{equation}
where the conformal factor $\Omega_o[g,\pi;x)$ satisfies the Lichnerowicz-York equation $8\Delta \Omega_o=\left(\frac 1 6 \langle \pi\rangle^2-2\Lambda\right)\,\Omega_o^5+R\,\Omega_o-\frac{\sigma^a_b\sigma^b_a}{|g|}\,\Omega_o^{-7}$, where $\sigma^a_b$ denotes the trace free part of the metric momenta and triangle brackets denote the mean taken w.r.t. $\sqrt{|g|}$. We see: (1) Shape Dynamics disentangles dynamics form constraints, (2) all constraints are linear in the momenta and generate geometric transformations. The price for this is the complicated form of the Hamiltonian $H_{SD}$. The equivalence of the ADM and the Shape Dynamics description is manifest when the ADM system is evolved in CMC-gauge $\pi(x)=\langle\pi\rangle\sqrt{|g|}(x)$. The initial value problem and the equations of motion of this system coincide with Shape Dynamics in the gauge $\Omega_o[g,\pi;x)-\langle\Omega^6_o[g,\pi]\rangle^{\frac 1 6}=0$. We will now consider how familiar bulk-gravity/boundary-CFT dualities can be seen as a consequence of 
the bulk-GR/bulk-SD equivalence.

\section{Asymptotically equivalent dynamics}

The familiar semi-classical holographic RG approach to the (A)dS/CFT correspondence can be discussed as a near boundary expansion of the general solution to Einstein's equations with particular asymptotic behavior\cite{ADS-refs}. Many of these asymptotic conditions can be translated into an ADM evolution with asymptotically homogeneous lapse and asymptotically vanishing shift. Moreover,  the Hamiltonian approach to holographic AdS/CFT  implies asymptotic homogeneity of $\pi/\sqrt{|g|}$ and spatial $R$. This implies that the lapse solves asymptotically the CMC-lapse equation:
\begin{equation}\label{equ:LFE}
 (\Delta_g-R-\frac{\langle\pi\rangle^2}{4})N=C,
\end{equation}
where the Laplacian $\Delta_g$ satisfies a maximum principle and $C$ is a constant. This means that the {\it{asymptotic boundary conditions imply the specific gauge on the ADM evolution that manifestly coincides with Shape Dynamics.}} The asymptotic conformal symmetry in $g_{ab}$ is thus the conformal symmetry of Shape Dynamics. However, the ADM evolution with homogeneous lapse does not propagate the CMC condition, which is why the {\it{manifest asymptotic equivalence of General Relativity with a line element of the form $ds^2=dt^2-g_{ab}\,dx^adx^b$ and Shape Dynamics is lost in the bulk.}}

\section{Explicit equivalence}

There are at least two gravity models in which the homogeneous lapse propagates the CMC gauge condition also in the bulk: strong gravity (spatial derivatives are neglected) and pure gravity on the 2+1-torus\cite{2+1-refs}. This is why for these two systems one can find manifest equivalence between the homogeneous lapse evolution and the evolution of a conformal theory  not only asymptotically but also in the bulk. This equivalence can be seen by putting the homogeneous lapse Hamiltonian for the strong gravity Hamiltonian
\begin{equation}
 S(N\equiv 1)=\int_\Sigma d^3x\left(\frac{1}{\sqrt{|g|}}\pi^{ab}(g_{ac}g_{bd}-\frac 1 2 g_{ab}g_{cd})\pi^{cd}+2\Lambda\sqrt{|g|} \right)\approx 0
\end{equation}
next to the Shape Dynamics volume constraint for this system
\begin{equation}
 \frac{\left(\int \sqrt{\sigma^a_b\sigma^b_a}\right)^2}V-\left(\frac 1 6 \langle\pi\rangle^2-2\Lambda\right)V\approx 0.
\end{equation}
One can explicitly check that the two Hamiltonians coincide when the inhomgeneous ADM constraints and the CMC-gauge conditions are satisfied. The Hamiltonians for pure gravity on the 2+1 torus are analogous; the homogeneous lapse Hamiltonian coincides with the generator of conformal dynamics on Teichm\"uller space.

A different way to see the manifest equivalence of Shape Dynamics with the homogeneous lapse evolution in the ADM system is by looking at the first orders of an asymptotic large volume expansion\cite{volume-ref} of the Shape Dynamics volume constraint
\begin{equation}
 0 \approx \left( 2\Lambda - \frac 1 6 \langle\pi\rangle^2 \right) - \left(\frac{V_0}{V}\right)^{2/3} \langle\tilde R\rangle + \left(\frac{V_0}{V}\right)^2 \int_\Sigma d^3x\frac{\sigma^a_b\sigma^b_a}{\sqrt{\tilde g}} + O\left((\frac{V}{V_0})^{-8/3}\right),
\end{equation}
where tilde denotes Yamabe gauge. The leading orders at large CMC volume coincide with the large volume expansion homogeneous lapse ADM Hamiltonian in Yamabe gauge, but the subleading terms, which are important for the bulk evolution, show explicit deviations.

\section{No Conformal Mode Problem and Holographic RG}

An important feature of Shape Dynamics is that $\pi/\sqrt{|g|}$ is constrained to be a time variable. The physical kinetic term of Shape Dynamics is thus, unlike the ADM kinetic term, non--negative; it does not have a conformal mode problem. This positivity and the fact that the Shape Dynamics volume constraint coincides asymptotically with the homogeneous lapse Hamiltonian allows one to interpret the dynamics generated by the kinetic term of $S(N\equiv 1)$ as the semiclassical approximation to the UV-limit of the exact renormalization group dynamics generated by a physical coarse--graining operator as it is e.g. used in Polchinski's equation. This is very similar of the interpretation of near boundary gravitational dynamics in holographic renormalization.

\section{Shape Dynamics Inspired Modified Gravity}

The use of the gauge/gravity duality as a definition of a gravity theory through a boundary CFT has been frequently suggested in the literature. In the shape dynamics framework, this would amount to taking Shape Dynamics as the fundamental description and General Relativity as an effective description. If Shape Dynamics is taken as the fundamental theory, then one would expect its Hamiltonian to be local and the effective spacetime description to be in general non--local, reversing the situation in which General Relativity is viewed as the fundamental description. This suggests the consideration of a particular class of gravity theories that can be derived from local Shape Dynamics Hamiltonians. The strong gravity limit combined with an expansion in spatial conformal invariants (ordered by the number of spatial derivatives) then suggests the study of completely conformal (modified) Shape Dynamics Hamiltonians of the form
\begin{equation}
 H_{mod}=\int_\Sigma d^3x \left(\sqrt{\sigma^a_b\sigma^b_a}+\alpha(\tau) \,CS(\Gamma)+...\right),
\end{equation}
where $CS(\Gamma)$ denotes the Chern-Simons functional of the spatial Christoffel symbol and the parenthesis stands for terms with more than three spatial derivatives.

\section{Conclusions}

It is the purpose of this contribution to show that the Shape Dynamics formulation of gravitational dynamics is a natural framework to address questions concerning gauge/gravity duality, because interesting questions can be answered with very simple Shape Dynamics arguments. The summary of our argument is as follows:
\begin{enumerate}
 \item Shape Dynamics is a gauge theory of spatial diffeomorphisms and spatial conformal transformations and its dynamics coincides with General Relativity. The equivalence is manifest if General Relativity is evolved in CMC gauge and Shape Dynamics is evolved in $\Omega_o[g,\pi;x)-\langle\Omega^6_o[g,\pi]\rangle^{\frac 1 6}=0$ gauge.
 \item The equivalence of the bulk evolution of Shape Dynamics with General Relativity implies a bulk/bulk duality between the General Relativity description and the conformal Shape Dynamics description. Familiar bulk-gravity/boundry-CFT dualities arise as a restriction of the description to the boundary.
 \item A simple spacetime description of bulk-gravity/boundary-CFT dualities through General Relativity uses a homogeneous lapse. This hides the more general bulk/bulk duality, because homogeneous lapse propagates CMC-gauge only in special situations.
 \item The near boundary regime can be characterized by large CMC volume, which is one of the situations, where the homogeneous lapse propagates CMC gauge. This is the reason why the simple General Relativity description of gravity finds the bulk/boundary duality but not the bulk/bulk equivalence.
 \item There are situations where the homogeneous lapse propagates the CMC condition, e.g. pure gravity on the 2+1 torus or the strong gravity approximation in higher dimensions. These cases allow one to find a bulk-gravity/bulk-CFT equivalence in the General Relativity description.
 \item The momentum conjugate to the conformal degree of freedom is constrained to be time in the Shape Dynamics description. The physical kinetic term of the gravity Hamiltonian is thus non--negative. This allows one to reinterpret the ADM kinetic term as the UV--limit of a coarse graining operator of an exact RG equation; an interpretation that is similar to holographic renormalization.
 \item Shape Dynamics suggests an interesting class of modified gravity theories.
\end{enumerate}
{\bf Acknowledgements:} This work was supported in part by the Government of Canada through NSERC.

\end{document}